# Point spread function modeling and images restoration for cone-beam CT[*]

ZHANG Hua(张华)[1], HUANG Kui-Dong (黄魁东)[2], SHI Yi-Kai (史仪凯)[1], XU Zhe (徐哲)[2]

[1] School of Mechanical Engineering, Northwestern Polytechnical University, Xi'an 710072, China
[2] Key Lab of Contemporary Design and Integrated Manufacturing Technology (Northwestern Polytechnical University), Ministry of Education, Xi'an 710072, China

**Abstract:** X-ray cone-beam computed tomography (CT) has the notable features such as high efficiency and precision, and is widely used in the fields of medical imaging and industrial non-destructive testing, but the inherent imaging degradation reduces the quality of CT images. Aimed at the problems of projection images degradation and restoration in cone-beam CT, a point spread function (PSF) modeling method is proposed firstly. The general PSF model of cone-beam CT is established, and based on it, the PSF under arbitrary scanning conditions can be calculated directly for projection images restoration without the additional measurement, which greatly improved the application convenience of cone-beam CT. Secondly, a projection images restoration algorithm based on pre-filtering and pre-segmentation is proposed, which can make the edge contours in projection images and slice images clearer after restoration, and control the noise in the equivalent level to the original images. Finally, the experiments verified the feasibility and effectiveness of the proposed methods.

## 1. Introduction

X-ray cone-beam CT uses plane array detector to capture the projection images of the tested object, and reconstructs continuous slice images by the corresponding algorithm from these images. Cone-beam CT has some advantages such as fast scanning speed, same spatial resolution in slice inner and between slices, high precision, and so on. Cone-beam CT has been widely used in medical imaging and industrial non-destructive testing [1, 2].

Flat panel detector (FPD) is the most widely used imaging component in cone-beam CT, in which X-ray is converted into visible light through the scintillator, and then visible light is converted into electrical signal by the photodiode [3]. Current FPD cannot avoid the phenomenon of visible light scattering from the process. Considering the effect of ray source focus size, there must be some degradation in the actual projection images. Point spread function (PSF) is an important indicator to characterize the degradation in imaging process, and there are several methods to measure it. Rectangular grating measurement [4] simplifies the preparation of the target, but the fabrication of fine slits greatly increases the test cost. Slits measurement [5, 6] is a simple sample preparation and high accuracy method, but requires a good parallelization of the slits and the pixel array of FPD. Steps measurement [7-9] has the differential operator in the calculation from the edge spread function to line spread function, which makes it very sensitive to noises, and the calculated modulation transfer function will be lower than the result of slits measurement. Pinhole measurement [10, 11] is widely used for focal spot diagnostic and PSF measurement in various ray sources, but the ideal pinhole imaging in actual systems is difficult to obtain.

Actual imaging systems are generally regarded as linear shift invariant models, which are usually used in the image restoration [12]. Current image restoration methods generally include the following types [13-15]: Frequency-domain methods, such as inverse filtering and Wiener filtering; Linear algebra restoration methods, such as constrained least squares and unconstrained least squares; Nonlinear restoration methods, such as maximum entropy restoration, genetic evolution, and neural network. Although there are many types of image restoration methods, most of them have some problems such as high algorithmic complexity and strict limitations, and the application results in cone-beam CT need further verification and improvement.

[*] Supported by the National Science and Technology Major Project of the Ministry of Industry and Information Technology of China (Grant No. 2012ZX04007021), the Young Scientists Fund of National Natural Science Foundation of China (Grant No. 51105315), the Natural Science Basic Research Program of Shaanxi Province of China (Grant No. 2013JM7003), and the Fundamental Research Funds for the Central Universities (Grant No. JC20120226 and 3102014KYJD022).
1) E-mail: zhhuang@nwpu.edu.cn
2) E-mail: kdhuang@nwpu.edu.cn

In this paper, we established the PSF model for the cone-beam CT imaging system based on FPD, and through restoring the projection images with PSF, the degradation caused by X-ray source focus size and FPD visible light scattering was reduced to improve the quality of projection images and slice images.

## 2. PSF modeling of cone-beam CT

For very small pinhole imaging in cone-beam CT, let ideal projection image of the pinhole is $f(x,y)$, PSF of FPD is $h_d(x,y)$, additive noise is $n(x,y)$, * stands for the convolution operator, the degraded projection image is $g(x,y)$ which can be expressed as:

$$g(x,y) = f(x,y) * h_d(x,y) + n(x,y) \tag{1}$$

Average a number of $g(x,y)$ to get $\bar{g}(x,y)$, and above equation can be written as:

$$\bar{g}(x,y) = f(x,y) * h_d(x,y) \tag{2}$$

And then:

$$H(u,v) = G(u,v) / F(u,v) \tag{3}$$

Where $G(u,v)$, $H(u,v)$, and $F(u,v)$ is the Fourier transform of $\bar{g}(x,y)$, $h_d(x,y)$, and $f(x,y)$, respectively.

Because $\bar{g}(x,y)$ is known, $h_d(x,y)$ can be calculated by image restoration after getting $f(x,y)$. $f(x,y)$ can be calculated according to the actual imaging [16]:

$$f(x,y) = M \times \frac{s(x,y)}{S} \tag{4}$$

Where $M$ is the sum of pixel gray of the pinhole region in $\bar{g}(x,y)$, $S$ is the pinhole sectional area, $s(x,y)$ is the area of the pixel $(x,y)$ covered by $f(x,y)$.

$h_d(x,y)$ obtained by the above method is discrete, we use three-dimensional Gaussian function to fit $h_d(x,y)$ to further reduce the calculation error ($a$ and $b$ are the fitting coefficients):

$$G(x,y) = a \cdot \exp(-\frac{x^2 + y^2}{b^2}) \tag{5}$$

In the above method, when the pinhole diameter increases, the effect of ray source focus size to the projection images becomes obvious. Let $h_s(x,y)$ be the system's PSF, $h_f(x,y)$ be the focus' PSF, it is generally considered that $h_s(x,y)$ is determined by both $h_d(x,y)$ and $h_f(x,y)$ [17], so:

$$\bar{g}(x,y) = f(x,y) * h_s(x,y) \tag{6a}$$

$$h_s(x,y) = h_d(x,y) * h_f(x,y) \tag{6b}$$

To get $h_s(x,y)$, beam stop plate (BSP) with pinholes is placed close to FPD in Ref. [16], and a multi-pinhole imaging method to measure and evaluate PSF based on the restoration quality of slice images is proposed. The slice image quality is significantly improved and comprehensively achieved the best state when the projection images are restored with the PSF measured by the pinhole with same size to ray source focus. PSF of cone-beam CT obtained by this method is under a specified scanning condition, but in fact, it is variant on different scanning conditions. If PSF must be measured once changing the scanning condition, the scanning workload would increase significantly, and meanwhile the hardware life would reduce.

In the case of fixed ray source focus in cone-beam CT, the point spread mainly depends on two scanning factors: Ray energy level and radiation dose. Ray energy level is controlled by the ray source voltage, and radiation dose is characterized by output image gray of FPD. In other words, scanning voltage and image gray determine the specific shape of PSF together. Taking into account the shape of three-dimensional Gaussian function is determined by the parameters $a$ and $b$, so long as building the relation $a$, $b$ with scanning voltage and image gray, respectively, the calculation model of PSF is established under any scanning conditions. Because when the focus size is constant, the degradation of projection images is associated with projection magnification ratio, we place BSP on the center of rotary table, which is consistent with the object measured in the actual scan, to get a more accurate PSF of cone-beam CT, as shown in Fig. 1.

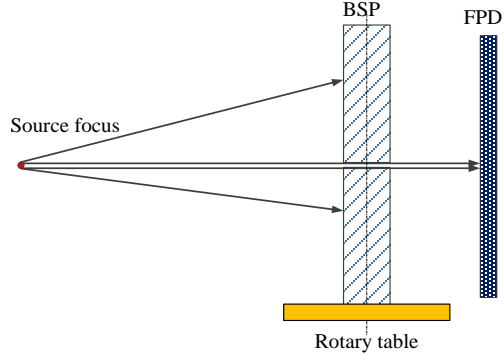

Fig. 1. Schematic illustration of PSF measurement with BSP.

Based on the above analysis and discussion, we propose a complete PSF modeling method for cone-beam CT as following steps:

(1) Fabricate a pinhole with same diameter to nominal size of the ray source focus on BSP;
(2) Get the PSF cross experimental parameters of scanning voltage and projection image gray as following steps:
　1) Select a number of arithmetic arranged scanning voltages within the cone-beam CT voltage range;
　2) Select a number of arithmetic arranged image grays within the FPD output gray range;
　3) For each scanning voltage, obtain the scanning current for each projection image gray without tested object and BSP.
(3) Get the center coordinates of pinhole projection with BSP [16];
(4) Obtain three-dimensional Gaussian PSF under each set of scanning parameters with PSF measurement method based on pinhole image restoration, as following steps:
　1) Using current scanning parameters, collect a number of projection images and average them to get $\bar{g}(x,y)$;
　2) According to Eq. (4), calculate the ideal image $f(x,y)$ of $\bar{g}(x,y)$;
　3) According to Eq. (3), get discrete $h_s(x,y)$ with the image restoration algorithm;
　4) Fit $h_s(x,y)$ with three-dimensional Gaussian function to get $h_1(x,y)$;
　5) Normalize $h_1(x,y)$ to get $h_2(x,y)$.
(5) Model PSF according to $h_2(x,y)$: Fit scanning voltage and projection image gray to the parameters $a$ and $b$, respectively, and then get two expressions which constitute the general PSF model of the cone-beam CT.

## 3. Projection images restoration of Cone-beam CT

Lucy-Richardson algorithm [18, 19] is an iterative nonlinear restoration algorithm based on Poisson distribution noise model, and its principle can be attributed to maximize the likelihood function, that is:

$$\max p(g(x,y)|f(x,y)) = \prod_{x,y} \frac{(h(x,y)*f(x,y))^{g(x,y)} \exp(-h(x,y)*f(x,y))}{g(x,y)!} \quad (7)$$

Take partial derivatives on $h(x,y)$, $f(x,y)$ in the above equation, respectively, and iterate by Picard to get:

$$h_{k+1}(x,y) = h_k(x,y)[f(-x,-y) * \frac{g(x,y)}{f(x,y)*h_k(x,y)}] \quad (8)$$

$$f_{k+1}(x,y) = f_k(x,y)[h(-x,-y) * \frac{g(x,y)}{h(x,y)*f_k(x,y)}] \quad (9)$$

Where $h_k(x,y)$ and $h_{k+1}(x,y)$ is the PSF of $k^{th}$ and $(k+1)^{th}$ iteration, respectively. $f_k(x,y)$ and $f_{k+1}(x,y)$ is the restored image of $k^{th}$ and $(k+1)^{th}$ iteration, respectively.

Eq. (8) and Eq. (9) indicate that $h(x,y)$ and $f(x,y)$ can be determined by cross iterative method to achieve blind image restoration, which is its special point. However, experimental results show that if there is no point spread estimation in the previous step, the restoration effect is poor. On the above discussion, we get $h_s(x, y)$ of cone-beam CT by Eq. (8), and in the next we can just use this feature to make projection images restoration by the same algorithm. Since the same algorithm is used for PSF calculation and images restoration, the calculation accuracy and reliability have certain advantages compared to the other images restoration algorithms. In the following we make some improvements from the noise control and computational efficiency for Lucy-Richardson algorithm used for a large number of projection images restoration in cone-beam CT.

Firstly, because the noises are not negligible in actual projection images, directly using of Lucy-Richardson algorithm will amplify the noises in the restored images, which is very unfavorable in observing image details. Taking into account the details and noises cannot be strictly distinguished for a certain extent in projection images, simple denoising will lost some image details inevitably. We take an approach called pre-filtering in this paper, that is: 1) Make the projection image denoising by non-local means algorithm based on GPU accelerated [20, 21] which has better overall performance; 2) Subtract denoised projection image from before to get the noise image; 3) Use Lucy-Richardson to restore denoised projection image; 4) Get the final restored image by adding the noise image to restored image. This method can not only keeps the approximate noise level in the images before and after restoration, but also retains as much as possible image details in the final restored images.

Secondly, the overall computation amount of Lucy-Richardson algorithm is relatively large. Considering that most of regions in projection images are usually blank and no tested object, these regions are not necessary to restore. In this paper, a reliable segmentation algorithm [22] is adopted to divide the smallest rectangle contained the tested object by costing less computation before the restoration, and make image restoration only for the rectangle region to greatly reduce the number of pixels to be restored. The step is called pre-segmentation, and is very necessary to rapidly restore the hundreds of projection images.

## 4. Experiments

### 4.1 PSF modeling experiment

The X-ray source of cone-beam CT used in the experiment is Y.TU 450-D02 of YXLON, FPD is PaxScan 2520 of Varian. BSP is a lead plate, which thickness is 20 mm, pinhole diameter is 2.5 mm and is same to the source focus size. $\overline{g}(x, y)$ is obtained by averaging 20 images. It should be noted that the center beam needs right through the pinhole of BSP in the experiment. Considering that the position of the center beam on FPD is known before cone-beam CT scanning [23], so adjust the position of BSP can meet the requirement.

The experiment designed and measured 16 PSFs under different sets of scanning voltage and image gray to build PSF model. Table 1 is the parameters $a$ and $b$ fitted by three-dimensional Gaussian function. The pinhole projection image under 200 kV and 2000 gray is shown in Fig. 2(a), corresponding normalized PSF is shown in Fig. 2(b).

Table 1. Fitted parameters of three-dimensional Gaussian function ($a/b$).

| Scanning voltage (kV) | Image gray | | | |
|---|---|---|---|---|
| | 1000 | 1500 | 2000 | 2500 |
| 100 | 0.02731/3.749 | 0.02713/3.765 | 0.02698/3.785 | 0.02684/3.798 |
| 200 | 0.02545/3.956 | 0.02508/4.005 | 0.02501/4.013 | 0.02491/4.019 |
| 300 | 0.02465/4.062 | 0.02455/4.078 | 0.02449/4.085 | 0.02442/4.095 |
| 400 | 0.02433/4.122 | 0.02425/4.134 | 0.02415/4.144 | 0.02408/4.157 |

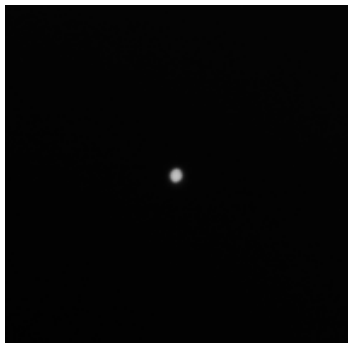 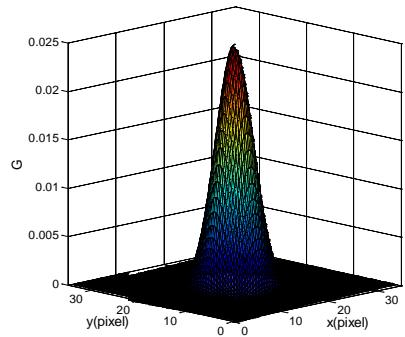

(a)          (b)

Fig. 2. PSF measurement: (a) pinhole projection (the display window is [0, 2000]), and (b) normalized PSF.

Using polynomial to fit the data in Table 1, we obtain Eq. (10) and Eq. (11), where $x$ is image gray and $y$ is scanning voltage. The three-dimensional shapes are shown in Fig. 3.

$$a = 0.03023 - 3.436 \times 10^{-7} x - 3.054 \times 10^{-5} y + 4.355 \times 10^{-10} xy + 4.056 \times 10^{-8} y^2 \quad (10)$$

$$b = 3.42 + 4.013 \times 10^{-5} x + 0.003341 y - 4.521 \times 10^{-8} xy - 4.105 \times 10^{-6} y^2 \quad (11)$$

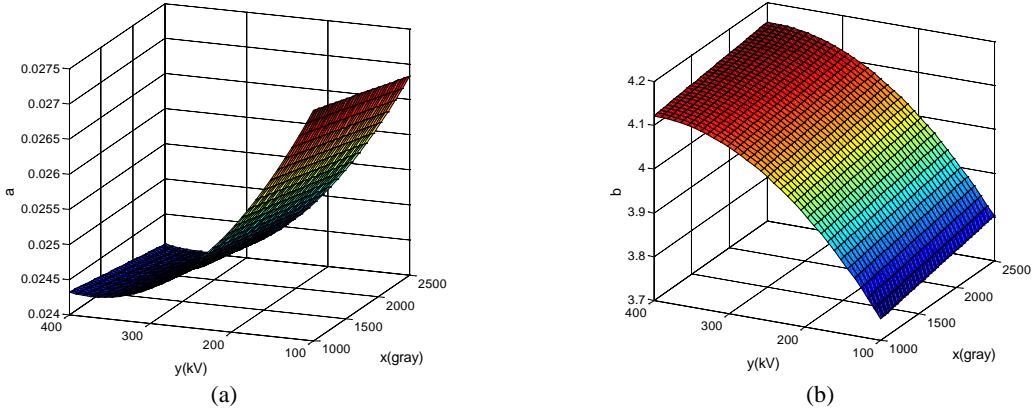

(a)          (b)

Fig. 3. Three-dimensional shapes of the fitting parameters: (a) $a$, and (b) $b$.

As can be seen from Table 1 and Fig. 3, the PSF of the experimental cone-beam CT is not changing large at different scanning parameters. Further investigating the PSF variation, we can draw the following three conclusions:

(1) When scanning voltage is fixed, the parameter $a$ ($b$) has approximately linear diminishing (increasing) relationship with image gray;

(2) When image gray is fixed, the parameter $a$ ($b$) has approximately square diminishing (increasing) relationship with scanning voltage;

(3) The impact of scanning voltage to the fitting parameters $a$ and $b$ is greater than that of image gray.

**4.2 Image restoration experiment**

An aluminum part is used to the projection images restoration experiment with the cone-beam CT, where scanning voltage is 180 kV, projection image gray is 2720, and the images collected are 360 with resolution of 1024×1024. The algorithms involved in comparison are Wiener filtering restoration and constrained least squares restoration. The numerical evaluation indicators include signal to noise ratio (SNR), contrast to noise ratio (CNR) and average gradient AG [24, 25]. The PSF under the scanning parameters can be calculated from Eq. (10) and Eq. (11):

$$G(x, y) = 0.02533 \cdot \exp(-\frac{x^2 + y^2}{3.975^2}) \quad (12)$$

Through the statistics of the restoration of 360 projection images, the total computation time (705 s) of our method including pre-filtering, pre-segmentation, and restoration, is only slightly more than that (654 s) of direct restoration for whole images. However, our method can control the noise within an acceptable level in restored images, which can be verified in the experimental data later. It should be noted that, although the pre-segmentation results are related to the size of parts, the actual tested parts are usually not bigger than that used in this experiment. So the computation time is representative.

Fig. 4 and Fig. 5 is the comparison before and after restoration of the first projection image and the 400[th] slice image (reconstructed by FDK algorithm), respectively. Where rectangle A is the comparison region of SNR, rectangle B is the comparison region of CNR and AG, line C is the comparison position of image gray. As can be seen from Fig. 4(e), our method make the edge of restored projection image sharper, outline clearer, noise keeping at about equivalent level with the original image, and there is no obvious ringing. Wiener filtering and constrained least squares both have a slight ringing. The quality performance of slice image is similar with that of projection image, which can be seen from Fig. 5(e). Furthermore, the ringing effect of Wiener filtering and constrained least squares in slice image is obviously stronger than that in projection image.

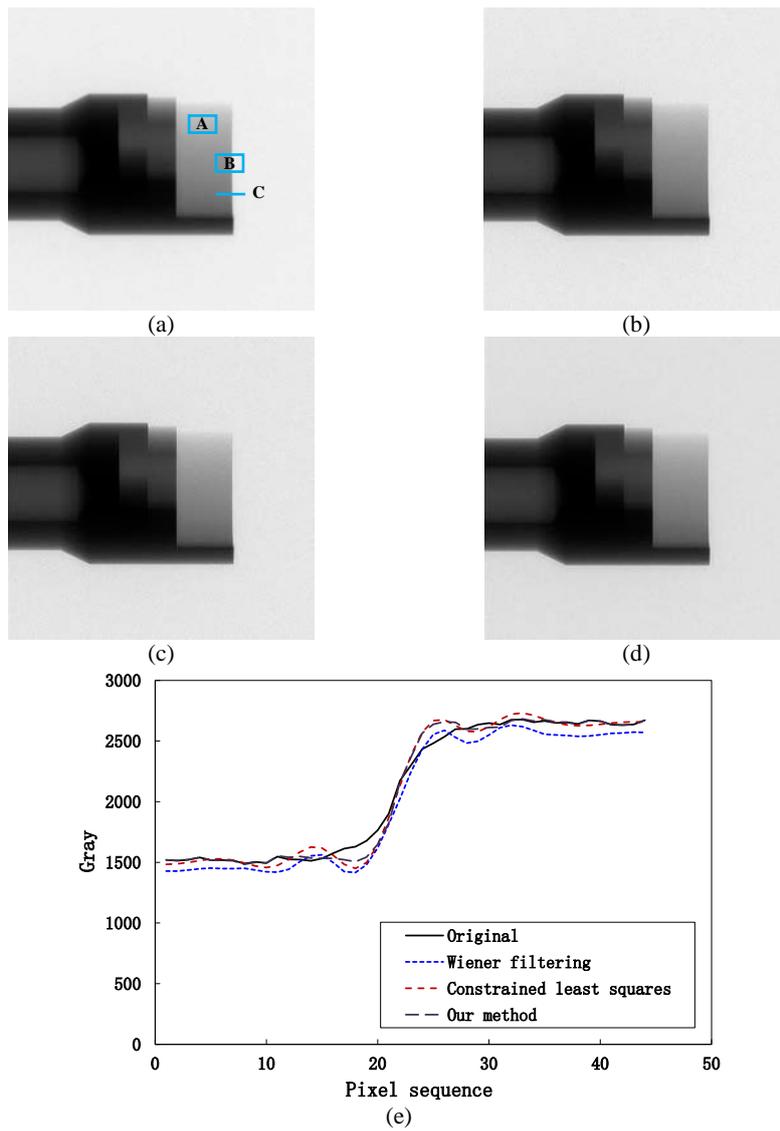

Fig. 4. Comparison before and after restoration of the first projection image: (a) original image, (b) Wiener filtering, (c) constrained least squares, (d) our method, and (e) gray profile. The display window is [0, 2720].

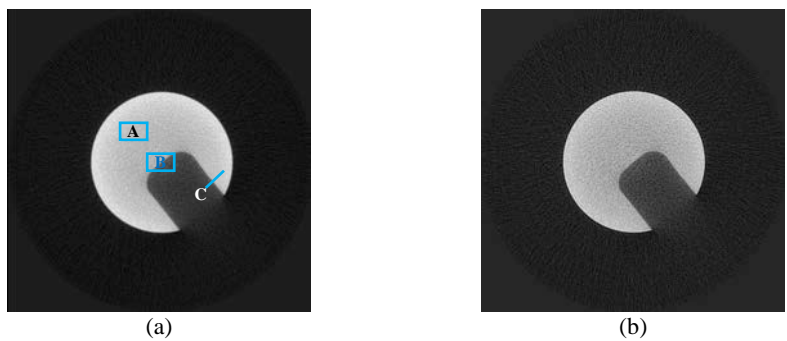

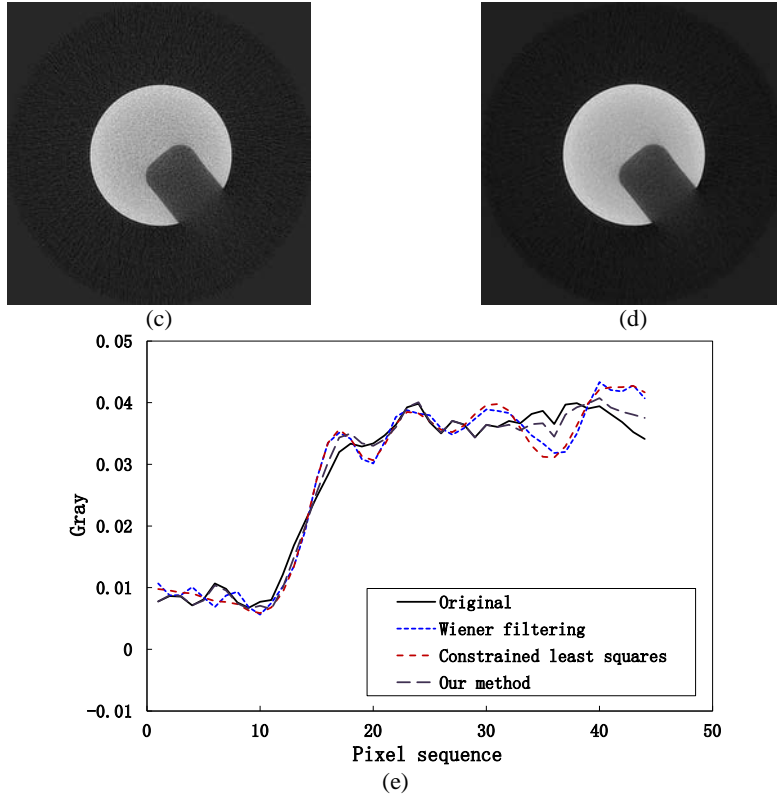

Fig. 5. Comparison before and after restoration of the 400th slice image: (a) original image, (b) Wiener filtering, (c) constrained least squares, (d) our method, and (e) gray profile. The display window is [-0.012, 0.068].

Table 2 is the numerical comparison of before and after restoration of the first projection image and the 400[th] slice image. We can see that the SNRs of projection and slice images restored with Wiener filtering and constrained least squares are significantly lower than that of the original, and our method can make it basically equivalent to the original, which shows that the noise has been effectively controlled in the image restoration process. In projection image, noise amplification makes the CNRs of Wiener filtering and constrained least squares smaller than that of the original, but our CNR is significantly higher than that of the original, which also benefited from the noise control. In slice image, because the restoration effect the CNRs of three methods are all higher than that of the original, but ours is the highest. Since AG cannot well distinguish details and noises, our method does not get the best in projection image, but get it in slice image. Overall, our method obtains the best quality in images restoration.

Table 2. Numerical comparison (the first projection image / the 400[th] slice image).

| Restoration method | Evaluation indicators | | |
| --- | --- | --- | --- |
| | SNR | CNR | AG |
| Original | 62.35/14.89 | 9.616/5.162 | 30.62/0.001648 |
| Wiener filtering | 56.30/11.96 | 9.097/5.260 | 36.11/0.001600 |
| Constrained least squares | 56.43/12.36 | 9.312/5.544 | 37.42/0.001483 |
| Our method | 61.63/14.95 | 10.81/5.556 | 36.06/0.001730 |

## 5. Conclusion

Based on the analysis and discussion of cone-beam CT imaging characteristics, we firstly proposed a complete PSF modeling method for cone-beam CT, and established a general PSF model of the imaging system. According to the model we can directly calculate the PSF under arbitrary scanning conditions without additional measurements, which improved the applying convenience greatly. Secondly, we proposed Lucy-Richardson restoration algorithm based on pre-filtering and pre-segmentation for projection images restoration, which can improve the clarity of projection images and slice images while controlling the noise in the equivalent level to the original images. The method can be directly applied to the X-ray digital radiography and cone-beam CT of fixed focus size. For the focus size changed significantly with the power, the impact of variable focus size to the general PSF model needs to be further evaluated.